\documentclass{iopart}

\begin{document}

\eqnobysec

\def\schw{Schwarzschild }

\newcommand{\preprint}[1]{\hfill{\sl preprint - #1}\par\bigskip\par\rm}

\def\dip{\smallskip Dipartimento di Fisica,
                                Universit\`a di Trento, Italia}
\def\infn{\smallskip Istituto Nazionale di Fisica Nucleare,\\
                                 Gruppo Collegato di Trento, Italia}
\def\dinfn{\smallskip Dipartimento di Fisica, Universit\`a di Trento\\
                           and Istituto Nazionale di Fisica Nucleare,\\
                                   Gruppo Collegato di Trento, Italia}
\def\Idip{\address{\dip}}
\def\Iinfn{\address{\infn}}
\def\Idinfn{\address{\dinfn}}
\def\dinbcn{\smallskip Unitat de Recerca, CSIC, IEEC, Edifici Nexus 104,\\
                                Gran Capit\`a 2-4, 08034 Barcelona, Spain\\
                   and Departament ECM and IFAE, Facultat de F\'{\i}sica, \\
              Universitat de Barcelona, Diagonal 647, 08028 Barcelona, Spain}
\def\Idinbcn{\address{\dinbcn}}

\def\eabs{\par\end{description}\hrule\par\medskip\rm}
\renewcommand{\date}[1]{\par\bigskip\par\sl\hfill #1\par\medskip\par\rm}

\def\M{{\cal M}}                       
\newcommand{\ca}[1]{{\cal #1}}         
\def\hs{\qquad}               
\def\nn{\nonumber}            
\def\beq{\begin{eqnarray}}    
\def\eeq{\end{eqnarray}}      
\def\ap{\left.}               
\def\at{\left(}               
\def\aq{\left[}               
\def\ag{\left\{}              
\def\cp{\right.}              
\def\ct{\right)}              
\def\cq{\right]}              
\def\cg{\right\}}             

\def\R{{\hbox{{\rm I}\kern-.2em\hbox{\rm R}}}}   
\def\H{{\hbox{{\rm I}\kern-.2em\hbox{\rm H}}}}   
\def\N{{\hbox{{\rm I}\kern-.2em\hbox{\rm N}}}}   
\def\C{{\ \hbox{{\rm I}\kern-.6em\hbox{\bf C}}}} 
\def\Z{{\hbox{{\rm Z}\kern-.4em\hbox{\rm Z}}}}   
\def\ii{\infty}                                  
\def\X{\times\,}                                  
\newcommand{\bin}[2]{\left(\begin{matrix}{#1\cr #2\cr}
                            \end{matrix}\right)}   

\def\Det{\mathop{\rm Det}\nolimits}                
\def\tr{\mathop{\rm tr}\nolimits}                  
\def\Tr{\mathop{\rm Tr}\nolimits}                  
\def\PP{\mathop{\rm PP}\nolimits}                  
\def\Res{\mathop{\rm Res}\nolimits}                
\def\res{\mathop{\rm res}\nolimits}                
\def\lap{\Delta}                                   
\def\cc{\phi_c}                                    
\def\ach{\cosh^{-1}}                               
\def\ash{\sinh^{-1}}                               
\def\ath{\tanh^{-1}}                               
\def\acth{\coth^{-1}}                              
\def\al{\alpha}
\def\be{\beta}
\def\ga{\gamma}
\def\de{\delta}
\def\ep{\varepsilon}
\def\ze{\zeta}
\def\io{\iota}
\def\ka{\kappa}
\def\la{\lambda}
\def\ro{\varrho}
\def\si{\sigma}
\def\om{\omega}
\def\ph{\varphi}
\def\th{\theta}
\def\te{\vartheta}
\def\up{\upsilon}
\def\Ga{\Gamma}
\def\De{\Delta}
\def\La{\Lambda}
\def\Si{\Sigma}
\def\Om{\Omega}
\def\Te{\Theta}
\def\Th{\Theta}
\def\Up{\Upsilon}


\title[Hawking Radiation as Tunneling]{Hawking Radiation as
  Tunneling: the $D$ dimensional rotating case}

\author{ M Nadalini$^1$,
L Vanzo$^1$ and S  Zerbini$^1$}
\address{$^1$ Department of Physics, Trento University and Gruppo
  Collegato INFN, \\ Trento, 38050 Povo, Italy  }
\ead{\mailto{nadalini@science.unitn.it},
  \mailto{vanzo@science.unitn.it}, \mailto{zerbini@science.unitn.it}}

\begin{abstract}
The tunneling method for the Hawking radiation is revisited and
applied to the $D$ dimensional rotating case. Emphasis is given to
covariance of results. Certain ambiguities afflicting the procedure
are resolved.
\end{abstract}

\section{Introduction}

It is well known that  black holes at classical level are black,
namely nothing
can escape from the event horizon, but if one takes into account the
quantum effects on the matter, they are no longer black
and the Hawking radiation is present. In particular, it has been shown
that this radiation is directly connected with the leading short
distance singularity of the two-point function of the matter fields on
the horizon\cite{Fredenhagen:1989kr}. Essentially this fact
can also be seen (with much less rigour, though) in the tunneling
method to be discussed below.

Is is also well known that there exist several derivations of
Hawking radiation and recently there has been a renewed interest
in a this long standing issue in QFT in curved space-time. In
fact, Parikh and Wilczek \cite{parikh00} introduced a
method involving the calculation of the classical action along
classically forbidden trajectories, starting just behind the
horizon and traveling outward to infinity. Thus, the particle
must travel backward in time, and its classical action  $I$
becomes imaginary (signaling the impossibility of such motion),
and this imaginary part can be viewed as a ``first quantization''
transition amplitude. The imaginary part is governed by the short
distance behavior on the horizon of the classical action. In their
tunneling approach, they  claimed the relevance of Painlev\'e
 stationary gauge  for 4 dim. Schwarzschild BH
\beq
ds^2=-(1-\frac{2mG}{r})dT^2-2mdrdT+dr^2+r^2dS^2_2\,. \nn
\eeq
Instead, we will show  how it is possible to work in the
$4$ dimensional  Schwarzschild  static gauge
\beq
ds^2=-(1-\frac{2MG}{r})dt^2+\frac{dr^2}{(1-\frac{2MG}{r})}+r^2dS^2_2\,, \nn
\eeq
and will  extend the discussion to $D$ dimensional rotating
cases.

The derivation is  quite elementary and  is based on the
 computation of
the  classical action $I$ along a trajectory
 starting from the horizon and ending in the bulk and the
associated   WKB approximation
\beq
\mbox{Amplitude} \propto e^{i I}\,.
\eeq
The related
 semiclassical emission rate is, ignoring the prefactor,
\beq
\Gamma \propto |\mbox{Amplitude}|^2 \propto
e^{-2\Im I}\,, \nn
\eeq
where $\Im I$ is the imaginary part of $I$. It will be shown that this
 semiclassical emission rate, can be written as a Boltzmann
 statistical factor 
\beq
\Gamma\propto e^{-\beta \,E}\,, \nn
\eeq
where  $E$ is the energy of the particle,
and $\beta$ will be interpreted  as inverse of temperature of the emitted
radiation.
\section{ The static case}

A generic static D-dimensional BH metric ansatz in the Schwarzschild
 static gauge, $x^\mu=(t, r ,x_i)$ reads
\beq
ds^{2} = -A(r)dt^2 + \frac{dr^2}{B(r)}+ C(r)h_{ij}dx^{i}dx^{j}=
g_{\mu \nu}dx^\mu dx^\nu \, .
\eeq
For black holes solutions, $A(r)$ and $B(r)$  vanish on the horizon $r=r_H$.
Beside the  WKB approximation,  the other  relevant  ingredient is the
 classical action $I$ of a particle in the above black hole
 background. One can evaluate it making use of the relativistic
 Hamilton--Jacobi equation
\beq
g^{\mu \nu}\partial_\mu I \partial_\nu I+m^2=0\, .
\eeq
Since the metric is static and spherical symmetric, one has
\beq
I=-E t+W(r)+J(x^i)\, ,
\eeq
and for $W(r)$, the following expression is obtained
\beq
\frac{dW}{dr}=\frac{1}{\sqrt{A(r)B(r)}}\sqrt{ E^2-A(r)\left(m^2+
\frac{ J^2}{C(r)}\right)}\,.
\eeq
The action along a path  connecting the horizon and a point in the
 bulk exterior region is
\beq
I=-Et+\!\int_{r_H}^{r_1}\!\frac{dr}{\sqrt{A(r)B(r)}}\sqrt{
 E^2-A(r)\!\left(m^2+\! 
\frac{ J^2}{C(r)}\right)}
+J(x_i)\,.
\eeq
For  non extremal black holes, the integrand has a non integrable
 singularity at the horizon which dominate the integral; splitting the
 integration into near-horizon and the bulk contributions, and considering only the  near
horizon one, we have \beq
A(r)=A'(r_H)(r-r_H)+\cdots\,,\,\,\,\,\,B(r)=B'(r_H)(r-r_H)+\cdots\,.
\eeq Thus, $\sqrt{A(r)B(r)}\propto (r-r_H)$.
 As a consequence, to define the integral we have to use  a
 regularization. However,  physical  Feynman prescription,
 naively  applied to the coordinate $r$, namely
\beq
\frac{1}{r-r_H} \rightarrow \frac{1}{r-r_H-i 0}=PP\frac{1}{r-r_H}
+i \pi \delta(r-r_H)
\eeq
turns to be   meaningless in curved space-times, due to the
 lack of covariance.

Let us illustrate this issue in $D=4$ and for Schwarzschild black hole
in the standard spherical coordinates $(t, r, \theta, \phi)$. Here we have
\beq
A(r)=B(r)=1-\frac{r_H}{r}\,,\,\,\,\ \,\,\,C(r)=r^2\,, \nn
\eeq
where $ r_H=2MG$ is the horizon radius. The classical action reads
\beq
I=-Et+\int_{r_H}^{r_1}  \frac{r dr}{r-r_H}\sqrt{ E^2-\frac{r-r_H}{r}\left(m^2+
\frac{ J^2}{r^2}\right)} +J(x_i) \,.
\eeq
The  naive use of the above prescription leads to
\beq
I=i\pi r_H E+\mbox{real part}=i\pi 2MG E+\mbox{real part}\,.
\eeq
For small $E$, we may, as leading approximation, neglect the back
reaction on the  black hole geometry. Thus
\beq
\Gamma \propto e^{-4 \pi MG\, E}\,,
\eeq
which turns to be  half of the correct result.

On the other hand, we  can use isotropic spherical coordinates,
 $r  \rightarrow  \rho$  defined by
\beq
r=\rho \at 1+\frac{\rho_H}{\rho}\ct^2\,,\,\,\,\rho_H=\frac{r_H}{4}\,.
\eeq
As a result, the general metric we started from,  reads with
 $\rho$ instead of $r$
\beq
ds^2=-A(\rho)dt^2 +\frac{1}{B(\rho)}\at d\rho^2+\rho^2dS_2^2\ct\,,  \nn
\eeq
where
\beq
A(\rho)=\at \frac{\rho-\rho_H}{\rho} \ct^2 \at
1+\frac{\rho_H}{\rho} \ct^{-2}\,, \nn
\eeq
\beq
B(\rho)=\at 1+\frac{\rho_H}{\rho} \ct^{-4}\,,\,\,\,
C(\rho)=\rho^2 \at 1+\frac{\rho_H}{\rho} \ct^4 \nn\,,
\eeq
\beq
\sqrt{A(\rho)B(\rho)}=
\frac{\rho-\rho_H}{\rho}\at 1+\frac{\rho_H}{\rho}\ct^{-3}\,. \nn
\eeq
The naive use of the prescription  in the previous formula leads to
\beq
I=i\pi 8 \rho_H E+\mbox{real part}=i\pi 4MG\, E+\mbox{real part}\,,
\eeq
namely  the correct result.
\beq
\Gamma \propto e^{-8 \pi MG \, E}\,,
\eeq
 since $T_H=\frac{1}{8\pi M G}$ is the Hawking temperature for the
Schwarzschild black hole.

 Why is present this discrepancy? Because the  covariance is missing!
 It can be recovered by means of   proper  spatial distance
\beq
d\sigma^2=\frac{dr^2}{B(r)}+C(r)h_{ij}dx^idx^j\,.
\eeq
This is a crono-invariant quantity, namely  invariant under the action
 of the group of time recalibration and
spatial diffeomorphisms \cite{Landau, Zelmanov, Cattaneo}.

Limiting to the s-wave
contribution (the bulk of particle emission is contained in such
modes), one gets
\beq
\sigma =\int\frac{dr}{\sqrt{B(r)}}\,,
\eeq
and
\beq
W(\sigma)=\int  \frac{d \sigma}{\sqrt{A(r(\sigma))}} \sqrt{
E^2-A(r(\sigma))m^2 }\,.
\eeq
Splitting again the near horizon and bulk contributions, in the
 so called near-horizon approximation,
\beq \sigma =\frac{2}{\sqrt{B'(r_H)}}\sqrt{r-r_H}+\cdots\,, \eeq
\beq A(r(\sigma))=A'(r_H)B'(r_H)\frac{\sigma^2}{4}+\cdots\,. \eeq
Thus, one gets the invariant result \beq
W(\sigma)=\frac{2}{\sqrt{A'(r_H)B'(r_H)}} \int  \frac{d
\sigma}{\sigma} \sqrt{ E^2-A(r(\sigma))m^2 }+\cdots\,. \eeq 

The function $1/\sigma$ restricted to positive $\sigma$ can only be
used to define a real distribution, with no imaginary part. Hence we
must let it to take a small negative value, the physical justification
for this being that we are discussing an off-shell process, whereby
the particle does not really travel in Lorentzian space-time during the 
tunneling process. 
 
Making then
use of the  Feynman prescription \beq \frac{1}{\sigma} \rightarrow
\frac{1}{\sigma-i 0}=PP\frac{1}{\sigma} +i \pi \delta(\sigma)\,,
\eeq it follows that $I$   acquires an imaginary part \beq
I=\frac{2\pi i}{\sqrt{A'(r_H)B'(r_H)}}\,E+ \mbox{(real
contribution)}\,. \eeq As a result, in the leading approximation
(small $E$ and no back reaction), the semi-classical emission rate
is given by the  general formula \cite{nada} \beq \Gamma \equiv
e^{-2 \Im I}= e^{-\frac{4\pi
E}{\sqrt{A'(r_H)B'(r_H)}}}=e^{-\beta_H E}\,. \label{g} \eeq This
is the standard Boltzmann factor as soon as one recognizes that
\beq \beta_H=\frac{4\pi}{\sqrt{A'(r_H)B'(r_H)}} \eeq represents
the inverse Hawking temperature of the radiation. This result is in
agreement with surface gravity and conical singularity methods.

The improved tunneling method also  correctly sets the
temperature for the  extremal black holes equal to zero \cite{nada}.

In these cases, the integral of the radial part of the action turns to be
divergent, but with a  pole of second order in $r-r_H$. This implies that
the horizon is at infinite proper distance, and   there is no
analytic covariant regularization giving an
imaginary part of the action.  As a consequence, the Hawking temperature is
 vanishing. We  stress that a naive use of the coordinate $r$
in computing and regularizing the integrals  leads to a non vanishing,
thus incorrect, temperature.

\section{The rotating case}
Our approach, with some reasonable approximation,  may be extended
to the rotating (stationary) case.

As a starting point, we consider a $D$ dimensional  black hole
solution in the Boyer-Lindquist coordinates $x^\mu=(t,r, x^a, x^A)$,
$x^a$ and $x^A$ being angular coordinates. We assume that
$g_{\mu\nu}=g_{\mu\nu}(r,x^A)$, and, as a consequence, $\partial_t
g_{\mu\nu}=0$ but also $\partial_a g_{\mu\nu}=0$. The angular
coordinate $x^a$ are associated with the axial symmetries which
are present and the only non vanishing off diagonal time-space
components are $g_{ta}$. Thus, we have \beq
ds^2=g_{tt}dt^2+2g_{ta}dtdx^a+g_{rr}dr^2+g_{ab} dx^a dx^b+ g_{AB}dx^A dx^B\,.
\eeq For our purpose, we are interested in the ADM-like form. It
reads
\beq ds^2&=&-N^2dt^2+g_{ab}(dx^a+ N^a dt)(dx^b+ N^b dt) \nn \\
&+&g_{rr}dr^2+g_{AB}dx^A dx^B\,,
\eeq where
\beq
A=-N^2=g_{tt}-(g_{ab})^{-1}g_{ta}g_{tb}\,,\,\,\,\,
N^a=-(g_{ab})^{-1}g_{ta}g_{tb}\,. \eeq This form of the metric
selects the two functions $A=-N^2$ and $B=g^{rr}=1/g_{rr}$. These two
functions are typically vanishing on the horizon and making use of
the near-horizon approximation,  one can formally deal with a
situation similar to the static cases previously considered. At
this point, one may recall that in the Kerr-Newman family of
solutions in-(out-)going geodesics are perpendicular to the
horizon, so the near horizon approximation implies the selection
of s-modes only, as far as the tunneling method is concerned.

Let us illustrate this general procedure with some examples.
The first non trivial  example is the $D=4$  Kerr-AdS black hole.
In  Boyer-Lindquist coordinates it reads
\beq
ds^2 &=& -{\Delta_r\over\rho^2}
\left[dt-\frac{a\sin^2\theta}{\Xi}\ d\phi\right]^2
+{\rho^2\over\Delta_r}\ dr^2+{\rho^2\over\Delta_\theta}\ d\theta^2\nn \\
&+&{\Delta_\theta\sin^2\theta\over\rho^2}
\left[a\ dt-\frac{r^2+a^2}{\Xi}\ d\phi\right]^2,
\label{KNAdS}
\eeq
where
\beq
\rho^2=r^2+a^2\cos^2\theta, \qquad \Xi=1-\frac{a^2}{ l^2},
\eeq
\beq
\Delta_r=(r^2+a^2)\at 1+\frac{r^2}{l^2}\ct-2mr , \qquad
\Delta_\theta=1-{a^2\over l^2}\cos^2\theta\,.
\eeq
Here $a$ denotes the rotational angular momentum parameter and the
negative cosmological constant is $\Lambda=-\frac{3}{l^2}$.

The ADM form, defining  the lapse function $A(r, \theta)$
and $g^{rr}=g_{rr}^{-1}= B(r, \theta)$ can be easily computed and reads
\beq
ds^2&=&-A(r,\theta)dt^2+\frac{dr^2}{B(r,\theta)}+
\frac{\rho^2}{\Delta_\theta} d\theta^2 \nn \\
&+& \frac{\Sigma^2 \sin^2
\theta}{\rho^2\Xi^2}\left(  d\phi-\omega  dt \right)^2\,,
\eeq
where
\beq
A(r,\theta)=\frac{\rho^2  \Delta_\theta \Delta_r}{\Sigma^2}\,,
\hs B(r,\theta)=\frac{\Delta_r}{\rho^2}\,,
\eeq
\beq
\Sigma^2=\Delta_\theta(r^2+a^2)^2-\Delta_r a^2 \sin^2 \theta\,,
\eeq
\beq
\omega=\frac{a\left(\Delta_\theta (r^2+a^2)-\Delta_r\right)\Xi}{\Sigma^2} \,.
\eeq
At the  horizon, $g^{rr}=0$, namely $\Delta_r(r_H)=0$.
Note the $A$ and $B$, functions of the metric, have  an explicit
dependence on the angular coordinate $\theta$.

The  strategy can be described as follows:  one has to expand the
quantities $A$ and $B$  near the horizon   only with respect to
$r$ \beq A(r,\theta)=A'(r_H,\theta)(r-r_H)+\cdots\,' \eeq \beq
B(r,\theta)=B'(r_H,\theta)(r-r_H)+\cdots\,. \eeq
As a result \beq
ds^2&=&-A'(r_H,\theta)(r-r_H)dt^2+\frac{dr^2}{B'(r_H,\theta)(r-r_H)}\nn \\
&+&
\frac{\rho^2(r_H,\theta)}{\Delta_\theta} d\theta^2
+
\frac{\Sigma^2(r_H,\theta) \sin^2 \theta}{\rho^2(r_H,\theta)\Xi^2}
  d\chi^2\,,
\label{lecca}
\eeq
where
\beq
\chi= \phi-\Omega  t\,\hs \Omega =\omega(r_H)\,.
\eeq
We also may consider the trajectories with $\theta$ and $\chi$
constants. Actually, it is a well known property of the Kerr black hole,
shared by the KAdS solution, that along geodesics in surfaces
$\theta=\theta_0$ constant, the combination $\phi-\Om t$ is finite on
the horizon, while both $\phi$ and $t$ diverge.
Then, making use of the ansatz
$I=-Et+J\phi+W(r,\theta)$, and noting that local change of variable
$\chi=\phi-\Omega t$,  with $\Omega =\omega(r_H)$,
transforms $E$ into $E-\Omega J$, and considering only  particles on
geodesics with $\theta=\theta_0$ constant, one arrives  at a
related formal near-horizon static metric.
Thus, the general formula Eq. (\ref{g}) gives
\beq
\Gamma \equiv e^{-2 \Im I}= \exp\left(-\frac{4\pi(E-\Omega
J)} {\sqrt{A'(r_H,\theta_0)B'(r_H,\theta_0)}}\right)\,.
\eeq
The dependence on $\theta_0$ is  only apparent, in fact an explicit
calculation leads to
\beq
T_H=\frac{3r_H^4+(l^2+a^2)r_H^2-a^2l^2}{4\pi r_H l(r_H^2+a^2)}\,,
\eeq
and this result is  in agreement with the one obtained by means of
surface gravity evaluated at the horizon.

As second example, let us consider the $ D=4+n $ dimensional rotating
 uncharged BH \cite{Myers 86},  having  only one non-zero angular moment.
This black hole solution may be phenomenologically relevant in the possible
BHs production at colliders within the brane world scenario, here $n$
denotes the number of extra dimensions. In fact
the colliding partons are supposed  to propagate on an infinitely-thin brane
 and therefore
they have a non-zero impact parameter only on a 2-dimensional plane along
the  brane. As a consequence   only one non-zero angular parameter
about an axis in the brane is present.
The associated metric induced on the brane  reads \cite{kanti04}
\beq
ds^2&=&-\left(1-\frac{\mu}{\rho^2\,r^{n-1}}\right)dt^2-
\frac{2 a\mu\sin^2\theta}
{\rho^2\,r^{n-1}}\,dt\,d\varphi+\frac{\rho^2}{\Delta}dr^2  \nn \\
&+&\rho^2\,d\theta^2+\left(r^2+a^2+\frac{a^2\mu\sin^2\theta}{\rho^2\,r^{n-1}}
\right)\sin^2\theta\,d\varphi^2, \nn
\eeq
where
\beq
\Delta=r^2+a^2-\frac{\mu}{r^{n-1}}\,, \quad
\quad\rho^2=r^2+a^2\cos^2\theta  \nonumber\,,
\eeq
with the mass and angular momentum
(transverse to the $(r, \varphi)$-plane\,) of the black hole
given by
\beq
M_{BH}=\frac{(n+2) A_{n+2}}{16 \pi G}\,\mu\,, \qquad
J=\frac{2}{n+2}\,M_{BH}\,a\,, \nonumber
\eeq
 $a$ being the angular momentum per unit mass, $G$ the
 $(4+n)$-dimensional Newton's
constant, and $A_{n+2}$ the area of a $(n+2)$-dimensional unit sphere given by
$A_{n+2}=2 \pi^{(n+3)/2}/\Gamma[(n+3)/2]$.

Again the formalism  requires the ADM   form of the metric, with
the knowledge of the lapse function $N^2=-A(r,\theta)$, and the  function
 $B(r,\theta)=g^{rr}$. The metric in the ADM form reads
\beq
ds^2&=&-\frac{\rho^2 \Delta}{(r^2+a^2)^2-\Delta a^2  \sin^2 \theta} dt^2+
\frac{\rho^2}{\Delta}\,dr^2\,+\rho^2\,d\theta^2  \nn \\
&+&\left(r^2+a^2+\frac{a^2\mu\sin^2\theta}{\rho^2\,r^{n-1}}\right)
\sin^2 \theta\, \left(  d\varphi -\omega dt \right)^2\,,
\eeq
where
\beq
\omega=\frac{a(r^2+a^2-\Delta)}{(r^2+a^2)^2-\Delta a^2  \sin^2 \theta}\,.
\eeq
Thus
\beq
A(r, \theta)=\frac{\rho^2 \Delta}{(r^2+a^2)^2-\Delta a^2  \sin^2 \theta}\,.
\eeq
As usual, the black hole horizon is defined by $g^{rr}=0$, namely
$\Delta(r_H)=0$.

Within the near horizon
approximation, one can repeat all the previous steps and the general formula
Eq. (\ref{g}) gives again
\beq
\Gamma \equiv e^{-2 \Im  I}= \exp\left(-\frac{4\pi(E-\Omega
J)} {\sqrt{A'(r_H,\theta_0)B'(r_H,\theta_0)}}\right)\,.
\eeq
The dependence on the fixed angle $\theta_0$ is again apparent and
 the Hawking temperature reads
\beq T_H=\frac{1}{4\pi
r_H}\frac{(n+1)r_H^2+(n-1)a^2}{r_H^2+a^2}\,. \nn \eeq Note the
dependence on the number of extra spatial dimensions ${n}$.

We conclude this Section with the following remark. In all rotating  cases,
one  must have {$E-\Omega J>0$. This can be easily
proved  as follows: the energy and angular momentum of a
particle with four-momentum $p^a$ are $E=-p^aK_a$ and
$J=p^a\tilde{K}_a$, respectively, where $K=\partial_t$ and
$\tilde{K}=\partial_{\phi}$ is the rotational Killing field.
But
the Killing field which is time-like everywhere (including the
ergosphere) is not $K^a$, but is instead  $\chi=K+\Omega\tilde{K}$.
Hence a particle (including those with negative energy inside the
ergosphere) can escape to infinity if and only if $p_a\chi^a<0$,
which gives the wanted inequality
\beq
p^a(K_a+\Omega\tilde{K}_a)=-E+\Omega J<0\,.
\eeq
At the same time, it is violated only in the super-radiant regime,
where the Boltzmann distribution must be replaced with the full
Planck distribution, and thus it is outside the validity of the
 semi-classical method.

\section{Conclusions}

 The tunneling method has been reformulated in the case of an
arbitrary static black
hole solution  and restricted only to the leading term, namely
neglecting the back reaction on the black hole geometry.
The inclusion of back reaction effects  can be done and gives rise to
sub leading correction in $E$ \cite{parikh00,vage}. The  novelty of our
 approach has mainly been consisted  in the covariant treatment of the
 horizon singularity, through the use of spatial proper distance. As a
 result, we have been able to derive the correct Hawking temperature,
 working in the static
Schwarzschild gauge.

The approach can be also extended to  extremal cases (BH solutions
having the horizon at infinite  proper spatial distance) obtaining
vanishing  Hawking temperature~\cite{nada}.

We have also considered  rotating  cases in the Boyer-Lindquist gauge.
 In particular the $4+n$ dimensional
rotating asymptotically flat BH with only one non-zero angular momentum.
 Again, making use
of the near horizon approximation, the tunneling rate in the
leading approximation has been derived and, as a consequence, an
expression of the Hawking temperatures  in agreement with the values
computed by means of evaluation of the surface gravity at horizon, has been
obtained. The approach can be easily extended to   generic
D dimensional rotating BH solutions that have recently been found
 \cite{Gibb2005}.


\end{document}